\documentclass[floatfix, noshowpacs, preprintnumbers, twocolumn, amsmath, amssymb, aps]{revtex4}

\usepackage{bm}
\usepackage{amsmath}
\usepackage{amsthm}
\usepackage{amssymb}
\usepackage{graphicx}
\usepackage{color}
\usepackage{bbold}
\usepackage{url}
\usepackage[caption=false]{subfig}

\newcommand{\bra}[1]{\left< #1 \right|} 
\newcommand{\ket}[1]{\left| #1 \right>}

\newtheorem{theorem}{Theorem}

\begin{document}
\title{Reliable computation from contextual correlations}

\author{Andr\'{e} L. Oestereich}
\affiliation{Instituto de F\'isica, Universidade Federal Fluminense, Av. Gal. Milton Tavares de Souza s/n, 
Gragoat\'a, Niter\'oi, RJ, 24210-340, Brazil}

\author{Ernesto F. Galv\~ao}
\affiliation{Instituto de F\'isica, Universidade Federal Fluminense, Av. Gal. Milton Tavares de Souza s/n, 
Gragoat\'a, Niter\'oi, RJ, 24210-340, Brazil}

\date{\today}

\begin{abstract}
An operational approach to the study of computation based on correlations considers black-boxes with one-bit inputs and outputs, controlled by a limited classical computer capable only of performing sums modulo-2.  In this setting, it was shown that non-contextual correlations do not provide any extra computational power, while contextual correlations were found to be necessary for the deterministic evaluation of non-linear Boolean functions. Here we investigate the requirements for \textit{reliable}  computation in this setting, that is, the evaluation of any Boolean function with success probability bounded away from $1/2$. We show that bipartite CHSH quantum correlations suffice for reliable computation. We also prove that an arbitrarily small violation of a multipartite GHZ non-contextuality inequality also suffices for reliable computation.
\end{abstract}

\maketitle

\section{Introduction}

The one-way model of measurement-based quantum computation (MBQC) relies on a sequence of single-qubit measurements made on highly entangled multi-qubit quantum states \cite{RaussendorfB01, GrossE07} . In MBQC, we need to have a classical computer to determine the adaptive bases to be measured during the computational process, but this computer needs not be universal; it is sufficient that it be able to compute modulo-2 sums. MBQC has been shown to be equivalent to a number of other universal models of quantum computation, despite the remarkable conceptual difference that in it, the dynamics is provided by the act of quantum measurement itself. This suggests the question: what specific characteristic of quantum correlations enable quantum computation in this setting?

To address this question, one may describe correlations in a black-box, model-independent setting, and set out to prove results about what can be computed with different types of correlations. This was the approach of Anders and Browne \cite{AndersB09}, elaborated on later by Raussendorf \cite{Raussendorf13}. They considered a scenario in which a number of black-boxes receive one bit each, outputting also a single bit. These inputs and outputs are provided by and received by a classical control computer. As in MBQC, the classical computer is limited to performing pre-determined sums modulo-2 of previous outputs, to decide other black-box inputs, and also the computation's final answer. It was proven that if the black boxes' inputs and outputs are correlated in a non-contextual way, then no extra computational power becomes available to the control computer \cite{Raussendorf13}. Contextual correlations provided by 3-qubit Greenberger-Horne-Zeilinger (GHZ) \cite{GreenbergerHZ89} states, on the other hand, enable deterministic, universal classical computation in this setting (as they allow for the evaluation of the universal two-bit NAND gate \cite{AndersB09}). Raussendorf generalized these results, proving bounds on the average success probability of computation of general Boolean functions when only non-contextual correlations are available \cite{Raussendorf13}.

In this paper we obtain two results regarding the use of contextual correlations for computation, also assuming a control computer capable of performing only modulo-2 sums. We are interested in achieving \textit{reliable} computation, defined as the capacity of evaluating arbitrary Boolean functions with a probability of success bounded away from $1/2$. First, we show that bipartite, quantum Clauser-Horne-Shimony-Holt (CHSH) \cite{ClauserHSH69} correlations are sufficient for reliable computation. Then we show that reliable computation is possible using non-contextual correlations violating a multipartite GHZ-type inequality by an arbitrarily small constant $\Delta>0$.

This paper is organized as follows. In section \ref{sec review} we review related work both on measurement-based quantum computation, and on relevant classical schemes for fault-tolerant computation. In section \ref{sec chsh} we prove quantum CHSH correlations are sufficient for reliable computation. In section \ref{sec small} we prove small violations of GHZ inequalities are also sufficient for reliable computation, wrapping up with discussion and some perspectives in section \ref{sec conclusion}.

\section{Review} \label{sec review}

In this section we review some of the literature relevant to our purposes. In sub-section \ref{sec detghz} we review ways to deterministically evaluate general Boolean functions using non-adaptive measurements on GHZ states. In sub-section \ref{sec context} we review some results of ref. \cite{Raussendorf13}, regarding the connection between contextuality and the MBQC evaluation of general Boolean functions. Finally, in sub-section \ref{sec reliable} we review a scheme for reliable classical computation, originally proposed by von Neumann \cite{vonNeumann56} and later refined by Hajek and Weller \cite{HajekW91} and Evans and Schulman \cite{EvansS03}.

\subsection{Computation using quantum correlations} \label{sec detghz}

Anders and Browne proposed \cite{AndersB09} the following framework to analyze, in a black-box scenario, how one may perform computation based on correlations. They considered a control computer of limited power, with the goal of computing a Boolean function $f$ on a $n$-bit string $\vec{b}= b_1, b_2, ... b_n$. The control computer is a $\oplus L$ (pronounced ``Parity L") computer \cite{Damm90, AaronsonG04}, capable only of evaluating sums modulo-2 of any subset of the input bits (and possibly the constant $1$). The $\oplus L$ computer can send one-bit inputs to a number of black-boxes, each of which also returns a single-bit output. It can pre- and post-process inputs and outputs, and must output $f(\vec{b})$. This restriction on the control computer is inspired by the one-way model of measurement-based quantum computation \cite{RaussendorfB01}, in which this control computer, aided by the correlations present in cluster states, enables universal quantum computation. Raussendorf called this class of MBQC ``l2-MBQC" for ``MBQC with mod-2 linear classical processing" \cite{Raussendorf13}, a terminology we adopt here.

Anders and Browne observed that quantum bipartite correlations maximally violating the CHSH inequality enable a probabilistic l2-MBQC evaluation of AND$(b_0, b_1)$ with a success probability $p_{CHSH}=\cos^2 (\pi/8) \approx 0.854$ \cite{AndersB09}. The protocol uses a maximally entangled two-qubit state $\ket{\Phi}=(\ket{00}+\ket{11})/\sqrt{2}$. The first qubit is measured in the $Z$ basis if $b_0=0$, and in the $X$ basis if $b_0=1$. The second qubit is measured in the $(Z+X)/\sqrt{2}$ basis if $b_1=0$ and in the $(X-Z)/\sqrt{2}$ basis if $b_1=1$. An outcome $+1$ codes for output $0$, and $-1$ codes for output $1$. It is then easy to check that for all inputs, with probability $p_{CHSH}=\cos^2 (\pi/8)$ the sum modulo-2 of the outputs will equal AND$(b_0, b_1)$. Anders and Browne also showed \cite{AndersB09} that three-qubit Greenberger-Horne-Zeilinger (GHZ) correlations enable deterministic classical computation, as they implement the universal 2-bit NAND gate.

In \cite{HobanCLB11}, Hoban \textit{et al}. proposed an alternative scheme that results in a non-adaptive, deterministic l2-MBQC evaluation of arbitrary Boolean functions on $n$ bits, using at most $2^n-1$ qubits in a multipartite GHZ state. For any given Boolean function, it is necessary to solve a system of linear equations to determine each qubit's required measurement basis. The computation's result will then be simply the sum modulo-2 of the outcomes at each qubit. This generalizes the particular result of Anders and Browne for a NAND gate, and will be useful to obtain one of our main results in section \ref{sec small}.

\subsection{Contextuality and measurement-based computation} \label{sec context}

In ref. \cite{Raussendorf13}, Raussendorf considered the l2-MBQC evaluation of general Boolean functions, and proved rather general results, in particular the following two theorems:

\begin{theorem}[Raussendorf \cite{Raussendorf13}]
Let $M$ be a l2-MBQC which deterministically evaluates a Boolean function. If the function is non-linear mod 2, then $M$ must draw on correlations which are strongly contextual (in the terminology used e.g. in \cite{AbramskyB11}).
\end{theorem}

As required to state Theorem 2 below, we define the distance $\nu(f)$ of a Boolean function $f$ to the closest linear Boolean function as the minimum number of inputs for which the output of $f$ differs from the output of any single linear Boolean function. As an example, the AND function of two bits has $\nu(\text{AND})=1$, as its closest linear approximation (the linear function that outputs constant $0$) has an output different from that of AND only for the single input 11.

\begin{theorem}[Raussendorf \cite{Raussendorf13}]
Let $M$ be a l2-MBQC that probabilistically evaluates a Boolean function $f$ on $k$ bits of input, with average error probability $\bar{e}_f$ (the average is done over the outputs corresponding to all possible inputs). Let $\nu(f)$ be the distance of $f$ to the closest linear Boolean function. Then if $\bar{e}_f <\nu(f)/2^k$, then $M$ must draw on contextual correlations.
\end{theorem}

With Theorem 1, Raussendorf completely characterized the set of functions which can be evaluated by non-contextual l2-MBQC as the set of linear Boolean functions. This, incidentally, explains why there is no classical measurement-based computational model: non-contextual correlations fail to extend the computational power of a $\oplus L$ computer.

Theorem 2 yields a number of non-contextuality inequalities, as each non-linear function $f$ on $k$ bits, with distance $\nu(f)$ to the closest linear function, yields a different one. Each such non-contextuality inequality consists of $2^k$ terms, each of which represents the probability of error for a non-contextual l2-MBQC evaluation of a specific $k$-bit input. Summing over the error of all inputs and dividing by $2^k$ we obtain the average error $\bar{e}^{NC}_f$ of any non-contextual l2-MBQC evaluation of $f$. Raussendorf's Theorem 2 can then be reframed as a non-contextuality inequality, namely $\bar{e}^{NC}_f \ge \nu(f)/2^k$. A l2-MBQC evaluation of $f$ with an average error that violates this inequality implies contextuality.

\subsection{Reliable classical computation} \label{sec reliable}

An important practical problem at the time when classical electronic computers were first developed was the control of errors. Von Neumann \cite{vonNeumann56}, in an influential article, obtained some of the first theoretical results regarding the possibility of reliable computation using unreliable computer circuit elements. Von Neumann considered the \textit{noisy circuit} model, composed of \textit{$\epsilon$-noisy} gates. An $\epsilon$-noisy, $k$-input gate computes a $k$-bit Boolean function, outputting the correct answer with probability $1-\epsilon$, and its negation with probability $\epsilon$, independently of the input. This input-independent error model, albeit unrealistic in many situations, can be used to obtain a number of results on fault-tolerant computation, some of which we review below.

If we fix the number $k$ of input bits, we are interested in the maximum value of $\epsilon$ for which it is possible to compute reliably. Our definition of reliable computation is that e.g. of refs. \cite{HajekW91, EvansS03}: there is a $\delta<1/2$ so that for every Boolean function there exists a circuit using $\epsilon$-noisy, $k$-input gates that results in a worst-input error of at most $\delta$. If reliable computation is possible, then repeating the computation $n$ times and taking a majority vote of the outputs results in a computational process with an error probability that decreases exponentially as $n$ increases.

The reliable computation scheme devised by von Neumann was further elaborated on by Hajek and Weller \cite{HajekW91} and Evans and Schulman \cite{EvansS03}, and works roughly as follows. At a given point in the computation, there are multiple copies of the noisy computation being done in parallel. The error rate is then reduced and brought close to a fixed value smaller than $1/2$ by using $k$-MAJ (majority) gates, which output the most common bit-value among the $k$-bit input (for odd $k$). After this error-reduction (and error uniformization) stage, some more (noisy) computation can be done on the parallel copies of the computational process, which increases the error within certain specified bounds. We then proceed with alternating error-correction and computational stages, until the output can be obtained.

Let us now review this reliable computational scheme in more detail. Note that in \cite{HajekW91, EvansS03}, it was assumed that all gates are $\epsilon$-noisy, with the same value of $\epsilon$. We will review the original results, and then make an observation, useful later on, about the situation when the different gates are allowed to have different error rates. Evans and Schulman proved the following

\begin{theorem}[Evans and Schulman \cite{EvansS03}]
For $k$ odd and 
\begin{equation}
\beta_k=\frac{1}{2}-\frac{2^{k-2}}{k \binom{k-1}{\frac{k-1}{2}}} \label{eq betak}
\end{equation}
there exists $\delta<1/2$ such that all Boolean functions can be reliably computed (with maximum error $\delta$) by a noisy circuit if and only if the $\epsilon$-noisy gates composing it have $\epsilon<\beta_k$.
\end{theorem}

The error-correction stages are performed with $\epsilon$-noisy $k$-MAJ gates, and was shown to be successful if and only if $\epsilon<\beta_k$ \cite{EvansS03}. When successful, the noisy $k$-MAJ gates bring the error arbitrarily close to a fixed point $\eta<1/2$. Once the error is sufficiently close to $\eta$, it is time to do computation on the multiple noisy copies of the memory. For this error-reduction stage to be successful, the initial error must be in the interval $(\eta, 1/2)$.

For the computational stage, Hajek and Weller's proposal uses a noisy version of the XNAND gate \cite{HajekW91}, whose truth table is shown in Table \ref{table:1}. As XNAND($b_0, b_1, b_1$) = NAND($b_0, b_1$), this gate is computationally useful as it can be applied to a noisy $b_0$ and two noisy copies of bit $b_1$, to output a noisy approximation of NAND ($b_0, b_1$). Hajek and Weller proved that if the 3 input bits have the same error, then $\mu$-noisy XNAND gates for any $\mu<1/2$ are sufficient to obtain a computational stage whose outputs are incorrect with probability $\delta<1/2$ \cite{HajekW91}, as required for the reliable computation scheme.

\begin{table}[h!]
\begin{center}
\begin{tabular}{ |c|c|c|c| } 
 \hline
 $b_0$ & $b_1$ & $b_2$ & output \\ 
 \hline
 0 & 0 & 0 & 1\\ \hline
 0 & 0 & 1 & 1\\ \hline
 0 & 1 & 0 & 0\\ \hline
 0 & 1 & 1 & 1\\ \hline
 1 & 0 & 0 & 1\\ \hline
 1 & 0 & 1 & 0\\ \hline
 1 & 1 & 0 & 0\\ \hline
 1 & 1 & 1 & 0\\ 
 \hline
\end{tabular}
\caption{Truth table for XNAND gate.}
\label{table:1}
\end{center}
\end{table}

By alternating computational stages with error-correction stages, it is possible to run sufficiently many copies of the noisy computation in parallel while keeping the error within correctable levels, effectively achieving reliable computation. In summary, for reliable computation using the scheme by Hajek and Weller \cite{HajekW91} and Evans and Schulman \cite{EvansS03}, it is sufficient to have: 
\begin{enumerate}
  \item $\epsilon$-noisy $k-$MAJ (majority) gates with $\epsilon<\beta_k$, (see eq. \ref{eq betak}), for any odd $k \ge 3$;
  \item $\mu$-noisy $3$-input XNAND gates with $\mu<1/2$.
\end{enumerate}

\section{Reliable computation from bipartite quantum correlations} \label{sec chsh}

Anders and Browne have shown that bipartite quantum correlations cannot be used to obtain a $l2$-MBQC process yielding an $\epsilon$-noisy AND gate with $\epsilon<\sin^2(\pi/8)\equiv 0.1464$ \cite{AndersB09}. Such a gate, if possible, would correspond to quantum correlations violating the Tsirelson bound \cite{Cirelson80} for the CHSH Bell scenario.

Despite the impossibility of deterministically evaluating all Boolean functions using bipartite quantum correlations, we make progress on this problem by proving that these correlations allow for reliable computation. For this, it is sufficient to show how we can obtain the two gates which are required for the reliable computation scheme we reviewed in section \ref{sec reliable} above.

The 3-bit MAJ function has a decomposition using 3 XOR gates and a single AND gate as follows:
\begin{equation}
\text{3-}MAJ(a,b,c)=((a \oplus b)AND(a \oplus c))\oplus a. \label{eq 3maj}
\end{equation}
In section \ref{sec detghz} above, we reviewed the l2-MBQC computation of the AND gate using bipartite CHSH quantum correlations proposed by Anders and Browne \cite{AndersB09}. In particular, we have seen that the error probability of this AND gate is independent of the input, and is equal to $\epsilon=\sin^2(\pi/8)\equiv 0.1464$. In the language of the reliable computation scheme using MAJ gates, this is an $\epsilon$-noisy AND gate.

Using formula (\ref{eq 3maj}), it is easy to check that a CHSH-based AND, together with the perfect XOR gates (available from the $\oplus L$ control computer), results in a 3-MAJ gate that is also $\epsilon$-noisy. Since $\epsilon=\sin^2(\pi/8)<\beta_3=1/6$ [see equation (\ref{eq betak})], we see that bipartite CHSH quantum correlations result in $l2$-MBQC implementation of a 3-MAJ gate with sufficiently small $\epsilon$ for successful error-correction in the reliable computation scheme of \cite{HajekW91, EvansS03}.

The 3-bit XNAND function has the following decomposition using a single AND gate:
\begin{equation}
XNAND(a,b_1,b_2)=((a \oplus b_1)AND(a \oplus b_1 \oplus b_2))\oplus a \oplus 1.
\end{equation}
Again, using an $\epsilon$-noisy AND gate from CHSH correlations [$\epsilon=\sin^2(\pi/8)$] and deterministic XOR gates available to the control computer results in a $\mu$-noisy XNAND gate with $\mu=\sin^2(\pi/8)<1/2$. Note, however, that using quantum correlations for implementing XNAND gates in this scheme is overkill, as it is sufficient to have $\mu<1/2$. As we show below, this can be done using non-contextual bipartite correlations.

To perform a $\mu$-noisy AND$(a,b)$ gate with $\mu=1/4$, we use non-contextual correlations corresponding to a uniform distribution over the four linear functions $l_1 = 0$, $l_2 = a$, $l_3 = b$, and $l_4 = a \oplus b \oplus 1$. It is easy to check that this convex combination of linear Boolean functions implements a $\mu$-noisy AND gate with $\mu=1/4$. Together with the perfect XOR gates this results in a $\mu$-noisy XNAND gate with $\mu=1/4$, and which uses non-contextual bipartite correlations only.

We have thus established that we can perform reliable computation using bipartite quantum CHSH correlations. Interestingly, even somewhat degraded CHSH correlations would suffice, as quantum correlations do not adhere tightly to the bound $\epsilon<\beta_3$. This prompted us to investigate alternative schemes using $\epsilon$-noisy gates obtainable from quantum correlations on a larger number $k>2$ of qubits, which we address in the next section.

\section{Reliable computation from arbitrarily small violation of a GHZ inequality} \label{sec small}

As we have seen in the previous section, the computational stages of the reliable computation scheme do not require any contextuality at all, while the error-correction stages rely on $\epsilon$-noisy, $k$-input majority gates, with $\epsilon<\beta_k$ [see equation (\ref{eq betak})]. In this section we show that these can be implemented by violating a suitable multipartite GHZ inequality by an arbitrarily small amount.

We are interested in non-contextuality GHZ inequalities associated with l2-MBQC evaluation of Boolean functions, as discussed in \cite{Raussendorf13} and described in our brief review in section \ref{sec context}. Here we will consider inequalities associated with the average error in any non-contextual l2-MBQC evaluation of the $k$-MAJ function \cite{Raussendorf13}, for varying $k$. As we discussed, these functions' non-linearity $\nu (\text{$k$-MAJ})$ will provide us with bounds on the minimum error incurred by non-contextual resources. This non-linearity has been proven to be, for odd $k$ \cite{DalaiMS06}:
\begin{equation}
\nu (\text{$k$-MAJ})=2^{k-1}- \binom{k-1}{\frac{k-1}{2}}
\end{equation}
We can use this result to obtain a lower bound for the average error (over all inputs) that any l2-MBQC evaluation of the $k$-MAJ function must have, if restricted to non-contextual correlations:
\begin{equation}
\bar{e}^{NC}_{\text{k-}MAJ} \ge \frac{\nu (\text{$k$-MAJ})}{2^k} = \frac{1}{2}-\frac{1}{2^k} \binom {k-1}{\frac{k-1}{2}}. \label{eq ekmaj}
\end{equation}
This is the family of GHZ-type non-contextuality inequalities we will concern us here.

Now note that the lower bound $\nu (\text{$k$-MAJ})/2^k > \beta_k$ [see eq. (\ref{eq betak})], which means that non-contextual correlations cannot result in $\epsilon$-noisy $k$-MAJ gates with error small enough to achieve reliable computation using the scheme we have reviewed. Indeed, if they did, we would be able to do non-contextual l2-MBQC evaluation of arbitrary functions with bounded error, which would contradict Raussendorf's Theorem 2 (see section \ref{sec context}) with regards to highly non-linear functions (for example, the so-called bent functions \cite{SasaoBbook09}).

It is also easy to check that $[\nu (\text{$k$-MAJ})/2^k-\beta_k]$ decreases monotonically with $k$, and goes to zero as $k \to \infty$.  This quantity is the gap between the lower bound for error in non-contextual l2-MBQC evaluation of the $k$-MAJ function, and the upper bound for the error sufficient for reliable computation. To achieve reliable computation with an arbitrarily small violation of GHZ inequality ($\ref{eq ekmaj}$), it is enough to choose large $k$ (so the gap is small), and design an $\epsilon$-noisy, slightly contextual $k$-MAJ gate bridging that gap.

For that purpose, let us recall section \ref{sec detghz}, where we reviewed the paper by Hoban \textit{et al.} \cite{HobanCLB11} describing a non-adaptive l2-MBQC protocol capable of evaluating arbitrary Boolean functions on $k$ bits using a quantum GHZ state of at most $2^k -1 $ qubits. In particular, it is possible to implement a deterministic $k$-MAJ gate, for any $k$, using a GHZ state of at most $2^k-1$ qubits.

Now consider the following convex combination of a GHZ state and the maximally mixed state of $2^k-1$ qubits:
\begin{equation}
\rho = (1-2\epsilon) \ket{GHZ}\bra{GHZ} +2\epsilon \frac{1}{(2^{2^k-1})} \mathbb{1}. \label{eq rho}
\end{equation}
Hoban \textit{et al.}'s protocol applied to the GHZ state results in a deterministic evaluation of the $k$-MAJ function; when applied to the maximally mixed state, it results in a uniformly random output, i.e. a $\epsilon$-noisy evaluation of the $k$-MAJ function with $\epsilon=1/2$. When applied to the mixture given in eq. (\ref{eq rho}) above, the protocols then gives us an $\epsilon$-noisy $k$-MAJ gate for any chosen $\epsilon \in (0,1/2)$.

To achieve reliable computation with an arbitrarily small violation $\Delta$ of inequality ($\ref{eq ekmaj}$), it is sufficient to pick $k$ large enough so that $\nu (\text{$k$-MAJ})/2^k- \beta_k < \Delta$. Then use Hoban \textit{et al.}'s protocol on state (\ref{eq rho}) as described above to obtain an $\epsilon$-noisy $k$-MAJ gate, choosing $\epsilon=\nu (\text{$k$-MAJ})/2^k - \Delta<\beta_k$.

\section{Discussion} \label{sec conclusion}

We have shown that in the bipartite scenario, CHSH correlations suffice for reliable computation, and in fact, even somewhat degraded CHSH correlations also suffice, provided a correlation-based AND gate can be implemented with error rate $e_{AND}<\beta_3=1/6$. We have also shown that an arbitrarily small violation of a GHZ-type inequality is also sufficient for reliable computation, provided we have a sufficiently large number of qubits in a GHZ state. A natural open question is whether \textit{any} amount of contextuality, for any number of parties, is sufficient for reliable computation.

Note also that the reliable computation scheme we have used employ a number of gates that increases exponentially with the input size $n$; are more efficient schemes possible?

Besides finding out the contextuality requirements for reliable classical computation, it would be interesting to investigate what is needed for quantum computational advantage. Some results along these lines were obtained in \cite{HowardWVE14, RaussendorfBDOB-V17}, where the universal quantum computation scheme using magic-state distillation \cite{BravyiK05} was shown to require violation of non-contextuality inequalities.

\textit{Acknowledgements.} We would like to thank Dan Browne for helpful correspondence, and for showing us the single-AND decomposition of 3-MAJ. We acknowledge financial support by the Instituto Nacional de Ci\^{e}ncia e Tecnologia de Informa\c{c}\~{a}o Qu\^{a}ntica (INCT-IQ/CNPq - Brazil).

\bibliographystyle{apsrev}

\begin{thebibliography}{19}
\expandafter\ifx\csname natexlab\endcsname\relax\def\natexlab#1{#1}\fi
\expandafter\ifx\csname bibnamefont\endcsname\relax
  \def\bibnamefont#1{#1}\fi
\expandafter\ifx\csname bibfnamefont\endcsname\relax
  \def\bibfnamefont#1{#1}\fi
\expandafter\ifx\csname citenamefont\endcsname\relax
  \def\citenamefont#1{#1}\fi
\expandafter\ifx\csname url\endcsname\relax
  \def\url#1{\texttt{#1}}\fi
\expandafter\ifx\csname urlprefix\endcsname\relax\def\urlprefix{URL }\fi
\providecommand{\bibinfo}[2]{#2}
\providecommand{\eprint}[2][]{\url{#2}}

\bibitem[{\citenamefont{Raussendorf and Briegel}(2001)}]{RaussendorfB01}
\bibinfo{author}{\bibfnamefont{R.}~\bibnamefont{Raussendorf}} \bibnamefont{and}
  \bibinfo{author}{\bibfnamefont{H.~J.} \bibnamefont{Briegel}},
  \bibinfo{journal}{Phys. Rev. Lett.} \textbf{\bibinfo{volume}{86}},
  \bibinfo{pages}{5188} (\bibinfo{year}{2001}).

\bibitem[{\citenamefont{Gross and Eisert}(2007)}]{GrossE07}
\bibinfo{author}{\bibfnamefont{D.}~\bibnamefont{Gross}} \bibnamefont{and}
  \bibinfo{author}{\bibfnamefont{J.}~\bibnamefont{Eisert}},
  \bibinfo{journal}{Phys. Rev. Lett.} \textbf{\bibinfo{volume}{98}}, 220503
  (\bibinfo{year}{2007}).

\bibitem[{\citenamefont{Anders and Browne}(2009)}]{AndersB09}
\bibinfo{author}{\bibfnamefont{J.}~\bibnamefont{Anders}} \bibnamefont{and}
  \bibinfo{author}{\bibfnamefont{D.~E.} \bibnamefont{Browne}},
  \bibinfo{journal}{Phys. Rev. Lett.} \textbf{\bibinfo{volume}{102}}, 050502
  (\bibinfo{year}{2009}).

\bibitem[{\citenamefont{Raussendorf}(2013)}]{Raussendorf13}
\bibinfo{author}{\bibfnamefont{R.}~\bibnamefont{Raussendorf}},
  \bibinfo{journal}{Physical Review A} \textbf{\bibinfo{volume}{88}},
  \bibinfo{pages}{022322} (\bibinfo{year}{2013}).

\bibitem[{\citenamefont{Greenberger et~al.}(1989)\citenamefont{Greenberger,
  Horne, and Zeilinger}}]{GreenbergerHZ89}
\bibinfo{author}{\bibfnamefont{D.~M.} \bibnamefont{Greenberger}},
  \bibinfo{author}{\bibfnamefont{M.~A.} \bibnamefont{Horne}}, \bibnamefont{and}
  \bibinfo{author}{\bibfnamefont{A.}~\bibnamefont{Zeilinger}}, in
  \emph{\bibinfo{booktitle}{Bell's Theorem, Quantum theory, and conceptions of
  the universe}}, edited by
  \bibinfo{editor}{\bibfnamefont{M.}~\bibnamefont{Kafatos}}
  (\bibinfo{publisher}{Kluwer Academic, Dordrecht}, \bibinfo{year}{1989}),
  p.~\bibinfo{pages}{69}.

\bibitem[{\citenamefont{Clauser et~al.}(1969)\citenamefont{Clauser, Horne,
  Shimony, and Holt}}]{ClauserHSH69}
\bibinfo{author}{\bibfnamefont{J.~F.} \bibnamefont{Clauser}},
  \bibinfo{author}{\bibfnamefont{M.~A.} \bibnamefont{Horne}},
  \bibinfo{author}{\bibfnamefont{A.}~\bibnamefont{Shimony}}, \bibnamefont{and}
  \bibinfo{author}{\bibfnamefont{R.~A.} \bibnamefont{Holt}},
  \bibinfo{journal}{Phys. Rev. Lett.} \textbf{\bibinfo{volume}{23}},
  \bibinfo{pages}{880} (\bibinfo{year}{1969}).

\bibitem[{\citenamefont{von Neumann}(1956)}]{vonNeumann56}
\bibinfo{author}{\bibfnamefont{J.}~\bibnamefont{von Neumann}}, in
  \emph{\bibinfo{booktitle}{Automata Studies}}, edited by
  \bibinfo{editor}{\bibfnamefont{C.~E.} \bibnamefont{Shannon}}
  \bibnamefont{and} \bibinfo{editor}{\bibfnamefont{J.}~\bibnamefont{McCarthy}}
  (\bibinfo{publisher}{Princeton University Press}, \bibinfo{year}{1956}).

\bibitem[{\citenamefont{Hajek and Weller}(1991)}]{HajekW91}
\bibinfo{author}{\bibfnamefont{B.}~\bibnamefont{Hajek}} \bibnamefont{and}
  \bibinfo{author}{\bibfnamefont{T.}~\bibnamefont{Weller}},
  \bibinfo{journal}{IEEE Trans. Inform. Theory} \textbf{\bibinfo{volume}{37}},
  \bibinfo{pages}{388} (\bibinfo{year}{1991}).

\bibitem[{\citenamefont{Evans and Schulman}(2003)}]{EvansS03}
\bibinfo{author}{\bibfnamefont{W.~S.} \bibnamefont{Evans}} \bibnamefont{and}
  \bibinfo{author}{\bibfnamefont{L.~J.} \bibnamefont{Schulman}},
  \bibinfo{journal}{IEEE Transactions on Information Theory}
  \textbf{\bibinfo{volume}{49}}, \bibinfo{pages}{3094} (\bibinfo{year}{2003}).

\bibitem[{\citenamefont{Damm}(1990)}]{Damm90}
\bibinfo{author}{\bibfnamefont{C.}~\bibnamefont{Damm}}, \bibinfo{journal}{Inf.
  Proc. Lett.} \textbf{\bibinfo{volume}{36}}, \bibinfo{pages}{247}
  (\bibinfo{year}{1990}).

\bibitem[{\citenamefont{Aaronson and Gottesman}(2004)}]{AaronsonG04}
\bibinfo{author}{\bibfnamefont{S.}~\bibnamefont{Aaronson}} \bibnamefont{and}
  \bibinfo{author}{\bibfnamefont{D.}~\bibnamefont{Gottesman}},
  \bibinfo{journal}{Phys. Rev. A} \textbf{\bibinfo{volume}{70}}, 052328
  (\bibinfo{year}{2004}).

\bibitem[{\citenamefont{Hoban et~al.}(2011)\citenamefont{Hoban, Campbell,
  Loukopoulos, and Browne}}]{HobanCLB11}
\bibinfo{author}{\bibfnamefont{M.~J.} \bibnamefont{Hoban}},
  \bibinfo{author}{\bibfnamefont{E.~T.} \bibnamefont{Campbell}},
  \bibinfo{author}{\bibfnamefont{K.}~\bibnamefont{Loukopoulos}},
  \bibnamefont{and} \bibinfo{author}{\bibfnamefont{D.~E.}
  \bibnamefont{Browne}}, \bibinfo{journal}{New Journal of Physics}
  \textbf{\bibinfo{volume}{13}}, \bibinfo{pages}{023014}
  (\bibinfo{year}{2011}).

\bibitem[{\citenamefont{Abramsky and Brandenburger}(2011)}]{AbramskyB11}
\bibinfo{author}{\bibfnamefont{S.}~\bibnamefont{Abramsky}} \bibnamefont{and}
  \bibinfo{author}{\bibfnamefont{A.}~\bibnamefont{Brandenburger}},
  \bibinfo{journal}{New Journal of Physics} \textbf{\bibinfo{volume}{13}},
  \bibinfo{pages}{113036} (\bibinfo{year}{2011}).

\bibitem[{\citenamefont{Cirelson}(1980)}]{Cirelson80}
\bibinfo{author}{\bibfnamefont{B.~S.} \bibnamefont{Cirelson}},
  \bibinfo{journal}{Letters in Mathematical Physics}
  \textbf{\bibinfo{volume}{4}}, \bibinfo{pages}{93} (\bibinfo{year}{1980}).

\bibitem[{\citenamefont{Dalai et~al.}(2006)\citenamefont{Dalai, Maitra, and
  Sarkar}}]{DalaiMS06}
\bibinfo{author}{\bibfnamefont{D.~K.} \bibnamefont{Dalai}},
  \bibinfo{author}{\bibfnamefont{S.}~\bibnamefont{Maitra}}, \bibnamefont{and}
  \bibinfo{author}{\bibfnamefont{S.}~\bibnamefont{Sarkar}},
  \bibinfo{journal}{Designs, Codes and Cryptography}
  \textbf{\bibinfo{volume}{40}}, \bibinfo{pages}{41} (\bibinfo{year}{2006}).

\bibitem[{\citenamefont{Sasao and Butler}(2009)}]{SasaoBbook09}
\bibinfo{author}{\bibfnamefont{T.}~\bibnamefont{Sasao}} \bibnamefont{and}
  \bibinfo{author}{\bibfnamefont{J.~T.} \bibnamefont{Butler}},
  \emph{\bibinfo{title}{Progress in Applications of Boolean Functions}}
  (\bibinfo{publisher}{Morgan and Claypool Publishers}, \bibinfo{year}{2009}).

\bibitem[{\citenamefont{Howard et~al.}(2014)\citenamefont{Howard, Wallman,
  Veitch, and Emerson}}]{HowardWVE14}
\bibinfo{author}{\bibfnamefont{M.}~\bibnamefont{Howard}},
  \bibinfo{author}{\bibfnamefont{J.}~\bibnamefont{Wallman}},
  \bibinfo{author}{\bibfnamefont{V.}~\bibnamefont{Veitch}}, \bibnamefont{and}
  \bibinfo{author}{\bibfnamefont{J.}~\bibnamefont{Emerson}},
  \bibinfo{journal}{Nature} \textbf{\bibinfo{volume}{510}},
  \bibinfo{pages}{351} (\bibinfo{year}{2014}).

\bibitem[{\citenamefont{Raussendorf et~al.}(2017)\citenamefont{Raussendorf,
  Browne, Delfosse, Okay, and Bermejo-Vega}}]{RaussendorfBDOB-V17}
\bibinfo{author}{\bibfnamefont{R.}~\bibnamefont{Raussendorf}},
  \bibinfo{author}{\bibfnamefont{D.~E.} \bibnamefont{Browne}},
  \bibinfo{author}{\bibfnamefont{N.}~\bibnamefont{Delfosse}},
  \bibinfo{author}{\bibfnamefont{C.}~\bibnamefont{Okay}}, \bibnamefont{and}
  \bibinfo{author}{\bibfnamefont{J.}~\bibnamefont{Bermejo-Vega}},
  \bibinfo{journal}{Phys. Rev. A} \textbf{\bibinfo{volume}{95}},
  \bibinfo{pages}{052334} (\bibinfo{year}{2017}).

\bibitem[{\citenamefont{Bravyi and Kitaev}(2005)}]{BravyiK05}
\bibinfo{author}{\bibfnamefont{S.}~\bibnamefont{Bravyi}} \bibnamefont{and}
  \bibinfo{author}{\bibfnamefont{A.}~\bibnamefont{Kitaev}},
  \bibinfo{journal}{Physical Review A} \textbf{\bibinfo{volume}{71}},
  \bibinfo{pages}{022316} (\bibinfo{year}{2005}).

\end{thebibliography}

\end{document}